\documentclass[%
 reprint,
superscriptaddress,
 amsmath,amssymb,
 aps,
prm,
]{revtex4-1}

\usepackage{graphicx}
\usepackage{epstopdf}
\usepackage{dcolumn}
\usepackage{bm}
\usepackage{gensymb}
\usepackage{upgreek}
\usepackage{hyperref}
\usepackage{tabularx}

\preprint{APS/123-QED}

\begin{document}

\title{Physical properties and electronic structure of single-crystal KCo$_2$As$_2$}

\author{D.J. Campbell}
\thanks{These authors contributed equally}
\affiliation{Maryland Quantum Materials Center, Department of Physics, University of Maryland, College Park, Maryland 20742, USA}
\author{B. Wilfong}
\thanks{These authors contributed equally}
\affiliation{Maryland Quantum Materials Center, Department of Physics, University of Maryland, College Park, Maryland 20742, USA}
\affiliation{Department of Chemistry, University of Maryland, College Park, Maryland 20742, USA}
\author{M.P. Zic}
\affiliation{Maryland Quantum Materials Center, Department of Physics, University of Maryland, College Park, Maryland 20742, USA}
\author{G. Levy}
\author{M.X. Na}
\affiliation{Department of Physics and Astronomy, University of British Columbia, Vancouver, British Columbia V6T 1Z1, Canada}
\affiliation{Quantum Matter Institute, University of British Columbia, Vancouver, British Columbia V6T 1Z4, Canada}
\author{T.M. Pedersen}
\author{S. Gorovikov}
\affiliation{Canadian Light Source, Saskatoon, Saskatchewan S7N 2V3, Canada}
\author{P.Y. Zavalij}
\affiliation{Department of Chemistry, University of Maryland, College Park, Maryland 20742, USA}
\author{S. Zhdanovich}
\affiliation{Department of Physics and Astronomy, University of British Columbia, Vancouver, British Columbia V6T 1Z1, Canada}
\affiliation{Quantum Matter Institute, University of British Columbia, Vancouver, British Columbia V6T 1Z4, Canada}
\author{A. Damascelli}
\affiliation{Department of Physics and Astronomy, University of British Columbia, Vancouver, British Columbia V6T 1Z1, Canada}
\affiliation{Quantum Matter Institute, University of British Columbia, Vancouver, British Columbia V6T 1Z4, Canada}
\affiliation{The Canadian Institute for Advanced Research, Toronto, Ontario M5G 1Z8, Canada.}
\author{E.E. Rodriguez}
\affiliation{Maryland Quantum Materials Center, Department of Physics, University of Maryland, College Park, Maryland 20742, USA}
\affiliation{Department of Chemistry, University of Maryland, College Park, Maryland 20742, USA}
\author{J. Paglione}
\affiliation{Maryland Quantum Materials Center, Department of Physics, University of Maryland, College Park, Maryland 20742, USA}
\affiliation{The Canadian Institute for Advanced Research, Toronto, Ontario M5G 1Z8, Canada.}

\date{\today}

\begin{abstract}

We present a method for producing high quality KCo$_2$As$_2$ crystals, stable in air and suitable for a variety of measurements. X-ray diffraction, magnetic susceptibility, electrical transport and heat capacity measurements confirm the high quality and an absence of long range magnetic order down to at least 2~K. Residual resistivity values approaching 0.25~$\mu\Omega$~cm are representative of the high quality and low impurity content, and a Sommerfeld coefficient $\gamma$~=~7.3~mJ/mol K$^2$ signifies weaker correlations than the Fe-based counterparts. Together with Hall effect measurements, angle-resolved photoemission experiments reveal a Fermi surface consisting of electron pockets at the center and corner of the Brillouin zone, in line with theoretical predictions and in contrast to the mixed carrier types of other pnictides with the ThCr$_2$Si$_2$ structure. A large, linear magnetoresistance of 200\% at 14~T, together with an observed linear and hyperbolic, rather than parabolic, band dispersions are unusual characteristics of this metallic compound and may indicate more complex underlying behavior.

\end{abstract}

\maketitle

\section{Introduction}

High-temperature unconventional superconductivity in the iron pnictides has led to numerous studies of isostructural materials incorporating other transition metals. While many non-iron-based pnictogen compounds do indeed superconduct, it is clear that the presence of iron (even if partially substituted) is necessary to achieve a high transition temperature ($T_c$). For example, $T_c$ in the Ni-based system Ba(Ni$_{1-x}$Co$_x$)$_2$As$_2$ system has a maximum just above 2~K \cite{EckbergBaNiCoAs}, but can exceed 35~K in Ba$_{1-x}$K$_x$Fe$_2$As$_2$ \cite{RotterBaKFeAs}. In most cases, metal pnictides require chemical substitution or pressure to suppress a competing antiferromagnetic/structural transition and stabilize a superconducting ground state. However, exceptions are found in materials containing an alkali element \textit{A}~=~K, Rb, or Cs, as the \textit{A}Fe$_2$As$_2$ \cite{SasmalAFe2As2, BukowskiRbFe2As2}, \textit{A}Cr$_3$As$_3$ \cite{MuKCr3As3}, \textit{A}$_2$Cr$_3$As$_3$ \cite{WangA2Cr3As3}, and \textit{A}$_2$Mo$_3$As$_3$ \cite{ChenCrMnSCs} parent compounds are all superconductors, with a maximum $T_c$ of about 11~K for Cs$_2$Mo$_3$As$_3$ \cite{ZhaoCs2Mo3As3}.

Tangential to the work done exploring superconductivity in the Fe-, Ni- and Cr-based pnictides, there has been a fair amount of study of the alkaline earth (AE) cobalt-based counterparts, including the $AE$Co$_{2}$As$_{2}$ series ($AE$~=~Ca, Sr, Ba), whose members crystallize in the same ThCr$_{2}$Si$_{2}$ structure as the iron- and nickel-based superconducting systems \cite{Pandey_SrCo2As2, Quirinale_CaCo2As2, Anand_BaCo2As2}. The Ba and Sr-containing $AE$Co$_{2}$As$_{2}$ compounds do not exhibit any magnetic order down to 2~K, but have been shown to exhibit ferromagnetic (FM) and antiferromagnetic (AFM) spin correlations, respectively, through NMR and neutron measurements \cite{Jayasekara_SrCo2As2_neutron, Wiecki_SrCo2As2_NMR, Ahilan_BaCo2As2_NMR,NakajimaTwoThirds}. SrCo$_2$As$_2$ in particular exhibits both FM and AFM fluctuations at 5~K, and their competition impedes the formation of long-range order \cite{Li_SrCo2As2_neutron, Li_SrCo2As2_neutron_II}. CaCo$_{2}$As$_{2}$, on the other hand, has an antiferromagnetic transition \cite{Quirinale_CaCo2As2} at 52~K. The difference lies in the fact that CaCo$_{2}$As$_{2}$ crystallizes in a collapsed ThCr$_{2}$Si$_{2}$ structure, with a much shorter \textit{c}-axis and therefore smaller separation between layers; it is also non-stoichiometric, with about 7\% Co vacancy in the CoAs layers. These two factors tweak the electronic structure and cause stronger interlayer interactions, which helps to stabilize $A$-type collinear antiferromagnetic order \cite{Sapkota_CaCo2As2_neutron}. Overall, the key difference with the presence of cobalt is the unit increase in electron filling \cite{Mao_ACo2As2}. Through many different works, this has been shown to move the system away from the electronic instabilities leading to superconductivity in the iron-based system and toward competing magnetic interactions \cite{SefatBaCo2As2, Cheng_CaCo2As2_spinflop, Zhang_CaCo2As2_torque, Anand_CaCo2As2, Pandey_SrCo2As2}. 

To that end, to push a Co-based system closer to the electronic structure and electron filling observed in the iron-based superconducting pnictides, the oxidation state of ions in the CoAs layers should be increased. In the case of KFe$_{2}$As$_{2}$, Fe has a nominal oxidation state of Fe$^{2.5+}$ ($d^{5.5}$). In order to lower the electron count of Co$^{2+}$ ($d^{7}$) we can replace the alkaline earth metal with an alkali metal to yield Co$^{2.5+}$ ($d^{6.5}$) and determine if it then behaves similarly to iron. KCo$_2$As$_2$ is a good choice for this; it is the only Co-based alkali pnictide that has been synthesized to date. However, the previous reports were of room temperature structural measurements \cite{RozsaKCo2As2}, or low temperature magnetization measurements \cite{AryalKCo2As2, ZinthKCo2As2} where unwanted phases in powder samples seem to have affected the data. Another reason to look at KCo$_2$As$_2$ is because KFe$_2$As$_2$ differs even from the other iron-based superconductors, as a result of strong electron-electron interactions that lead to heavy fermion-like behavior at low temperatures \cite{HardyKFe2As2}. We have found a reliable method for producing KCo$_2$As$_2$ single crystals with millimeter-scale dimensions that are stable in air. As a result, we have been able to measure magnetic, electrical, and thermodynamic properties at low temperatures, as well as directly image the band structure through angle-resolved photoemission spectroscopy (ARPES). While there is no long range magnetic order, and electrical transport measurements indicate a lack of superconductivy above 100~mK, a linear magnetoresistance is observed from low fields up to 50~K, and photoemission experiments reveal unusual shapes in the band dispersions. Our measurements offer a comparison to the iron-based materials that was previously restricted to theory \cite{SinghKCo2As2}, and contribute to the understanding of the relationship between electron distribution, magnetism, and superconductivity in pnictide materials.

\section{Crystal Growth and Characterization}

\begin{table}[t]
    \centering
\caption{Single crystal x-ray diffraction data for KCo$_{2}$As$_{2}$.} 
\begin{tabular}{lcc}
\hline\hline
T (K)                         & 150(2)               & 301(2)           \\
Space group                           & $I4/mmm$             & $I4/mmm$             \\
Crystal system                        & tetragonal           & tetragonal           \\
$a$ (\AA)                       & 3.809(2)             & 3.805(2)             \\
$c$ (\AA)                       & 13.580(8)            & 13.573(8)            \\
Volume (\AA$^{3}$)                    & 197.0(2)             & 196.5(3)             \\
Z                                     & 2                    & 2                    \\
Calculated density (g/cm$^{3}$)       & 5.172                & 5.185                \\
$\lambda$, Mo K$\alpha$ (\AA)         & 0.71073              & 0.71073              \\
No. reflections collected             & 2307                 & 2220                 \\
No. independent reflections           & 184                  & 184                  \\
$F(000)$                              & 278.0                & 278.0                \\
$R_{1}, wR_{2}$                       & 0.0285, 0.0552       & 0.0264, 0.0538     \\ \hline\hline
\end{tabular}
\label{XRDBothTemps}
\end{table}

\begin{table*}[t]
	\centering	
\caption{Structural, lattice, and anisotropic displacement parameters for $I4/mmm$ KCo$_{2}$As$_2$ from single crystal data at 150~K. All off-diagonal terms are equal to zero.} 
\begin{tabular}{ l c c c c c r } 
\hline \hline									
Atom&\hspace{0.5cm}		Wyckoff Site\hspace{0.5cm}		&\hspace{0.5cm}	$x$\hspace{0.95cm}		&\hspace{0.95cm}		$y$\hspace{0.95cm}		&\hspace{0.95cm}	$z$ \hspace{0.75cm}		&\hspace{0.75cm}		$U_{11}$({\AA$^{2}$}) = $U_{22}$({\AA$^{2}$})\hspace{0.75cm}		&\hspace{0.75cm}	$U_{33}$({\AA$^{2}$})		\vspace{1mm} \\
\hline																		\\
K				&	2a		&	0			&		0			&	0.5					&			0.0115(5)		&	0.0199(9)			\\ 
Co				&	4d		&	0.5			&		0.5			&	0.75					&			0.0065(2)		&	0.0176(4)     			\\
As				&	4e		&	0.5			&		0.5			&	0.65040(5)				&			0.00728(19)		&	0.0171(3)      		\\ \hline\hline
											
\end{tabular}
\label{XRD150K}
\end{table*}

\begin{figure}[t]
    \centering
    \includegraphics[width=0.48\textwidth]{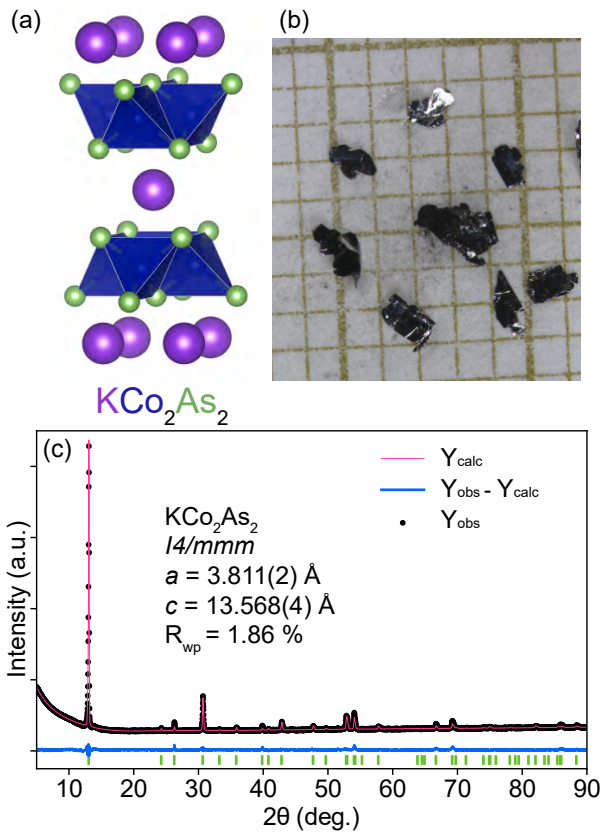}
    \caption{(a) Crystal structure of KCo$_{2}$As$_{2}$ showing two-dimensional CoAs planes with K$^{+}$ ions between adjacent layers. (b) Representative shiny, silver and thin crystals (on 1~$\times$~1 mm$^2$ paper) grown from the discussed method after remaining flux had been removed by washing with anhydrous ethanol. (c) Powder x-ray diffraction patterns with Cu K$\alpha$ radiation of KCo$_{2}$As$_{2}$ in the body-centered tetragonal structure (\textit{I4/mmm}) at room temperature (about 295~K). Green tick marks representing the corresponding tetragonal phase are shown below the calculated, observed, and difference curves from Rietveld analysis.}
    \label{fig:XRD}
\end{figure}

Crystals were grown from prereacted CoAs (Co powder, Alfa Aesar, 99.8\%; As lump, Alfa Aesar, 99.999\%) and K (cubes, Sigma Aldrich, 99.5\%) in a 2:3 ratio. The excess K acts as a flux at high temperatures, similar to what has been done for KFe$_2$As$_2$ \cite{KihouKFe2As2}. The combination was placed in an alumina crucible and sealed inside a stainless steel (grade 316 austenitic) pipe in argon atmosphere. The mixture of CoAs and K was heated to 900~$\degree{}$C at 50~$\degree{}$C/hour and held at that temperature for 12~hours to achieve a congruent melt, before being cooled to 650~$\degree{}$C at 2~$\degree{}$C/hr, at which point the temperature control was turned off and the furnace was allowed to cool to room temperature (roughly 20~$\degree{}$C). The reaction vessel was then cut open in an argon glove box and the growth material harvested from the alumina crucibles. This process resulted in a mixture of polycrystalline material and single crystals. The latter had dimensions that could exceed 1~$\times$~1~$\times$~0.5~mm$^3$ [Fig.~\ref{fig:XRD}(c)], with the shortest axis universally being the \textit{c}-axis, typical for layered tetragonal systems.

Despite previous reports of the material being moisture sensitive in powder form \cite{AryalKCo2As2, ZinthKCo2As2}, we found our KCo$_2$As$_2$ crystals to be air stable. After being left in ambient conditions for several days, there was no noticeable change in the x-ray diffraction (XRD) pattern, electrical transport results, or appearance of samples. However, the powder obtained directly from growths can decompose within an hour to CoAs and even ignite in air, likely because of unreacted K in the remaining flux. Thorough washing with anhydrous ethanol removes excess potassium arsenide fluxes and the remaining material is air stable. Similarly, crystals can have residual flux on their surface that will absorb water and decompose, but which can readily be removed with anhydrous solvents.

X-ray measurements were conducted on individual single crystals and a powder obtained from single crystals. Single crystal diffraction was performed on a Bruker Smart Apex II CCD diffractometer at both 300~K and 150~K. The integral intensity was corrected for absorption with the SADABS software \cite{KrauseSADABS} using the multi-scan method. The structure was solved with the ShelXT-2014 \cite{SheldrickSHELXT} program and refined with the ShelXL-2015 \cite{SheldrickSHELXL} program and least-square minimisation using ShelX software package \cite{SheldrickSHELXL}. Powder diffraction was done with a Bruker D8 x-ray diffractometer with Cu K$_{\alpha}$ radiation, $\lambda = 1.5406 $~\AA{} (step size = 0.020$\degree$). Rietveld refinements on powder diffraction data were carried out using the TOPAS 4.2 software \cite{TOPAS}. 

Single crystal XRD of one of the recovered samples verified that KCo$_2$As$_2$ forms in the ThCr$_2$Si$_2$ structure expected from previous reports \cite{RozsaKCo2As2, AryalKCo2As2, ZinthKCo2As2}, the same one as the ``122'' iron pnictides \cite{PaglioneFeSCs}, illustrated in Fig.~\ref{fig:XRD}(b). Lattice parameters were \textit{a}~=~3.805(2)~\AA{} and \textit{c}~=~13.573(8)~\AA{} at 300~K, and \textit{a}~=~3.809(2)~\AA{} and \textit{c}~=~13.580(8)~\AA{} at 150~K. Further details for both data sets are given in Table~\ref{XRDBothTemps}, and more extensive parameters of the 150~K refinement are in Table~\ref{XRD150K}. The 300 K lattice parameters are similar to those found previously \cite{RozsaKCo2As2, AryalKCo2As2}. We note also that the \textit{c}-axis length and $c/a$-axis ratio are both roughly similar to KFe$_2$As$_2$ \cite{RozsaKCo2As2} and BaCo$_2$As$_2$ \cite{AryalKCo2As2}, substantially higher than those of CaCo$_2$As$_2$ \cite{Anand_CaCo2As2}, i.e., the material does not form in the ``collapsed'' tetragonal structure. The slight negative thermal expansion (or, at least, the lack of positive thermal expansion) could be associated with the presence of Co and subtle changes in magnetism, as was noted for the spin glass CoSe \cite{WilfongCoSe}. Refinement of the single crystal pattern verified full elemental occupation of all structural sites. Room temperature (about 295~K) powder diffraction gave similar values of \textit{a}~=~3.8102(6)~\AA{} and \textit{c}~=~13.563(3)~\AA{}, and the powder pattern can be seen in Fig.~\ref{fig:XRD}(a) together with the refined fit.

\section{Magnetic Properties}

Previous magnetic measurements of KCo$_2$As$_2$ performed on powder samples \cite{AryalKCo2As2, ZinthKCo2As2} yielded contradictory results: one study reported a ferromagnetic transition \cite{AryalKCo2As2} at 15~K, while the other found no long range ordering \cite{ZinthKCo2As2}. Band structure calculations predicted a lack of order in this system, and an overall weaker magnetic response than KFe$_2$As$_2$ \cite{SinghKCo2As2}. This would be consistent with what we can infer from being in the uncollapsed tetragonal state, which in Co-based 122s typically inhibits long range order \cite{Quirinale_CaCo2As2}. We indeed found that crystals had a very small magnetic susceptibility $\chi$, about an order of magnitude lower than that of KFe$_2$As$_2$ \cite{ChengPnictides}. To increase the signal, measurements were made on a collection of four single crystals with a total mass of 1.7~mg, coaligned so that field was parallel to the $ab$ plane. Samples were measured using a 7~T Quantum Design MPMS3 system attached with GE varnish to a quartz paddle.

\begin{figure}
    \centering
    \includegraphics[width=0.48\textwidth]{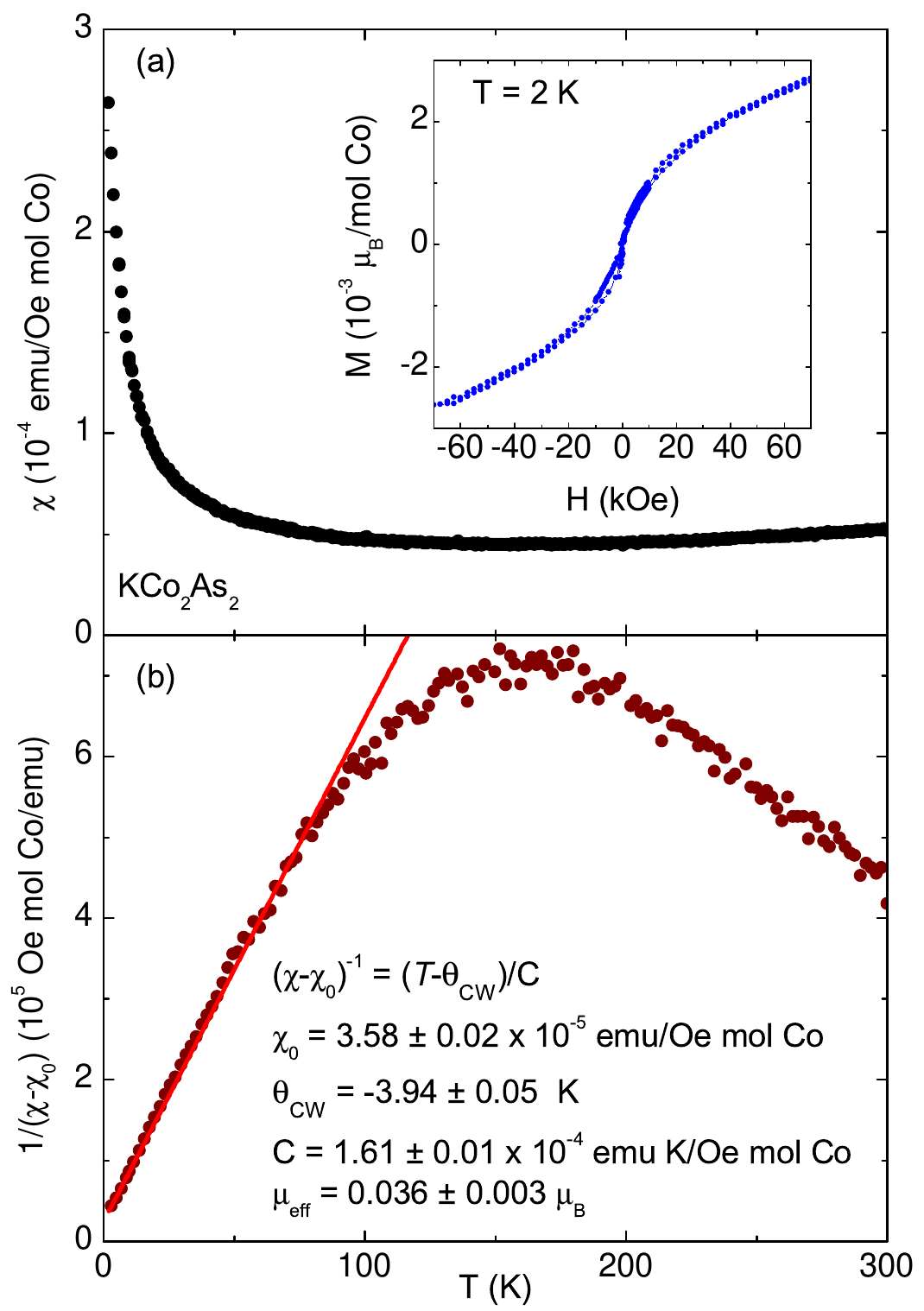}
    \caption{(a) Magnetic susceptibility of several coaligned KCo$_2$As$_2$ single crystals as a function of temperature, with 1~T field applied parallel to the \textit{ab}-plane of the crystals. The inset shows the magnetization as a function of field at base temperature during the same measurement. (b) Inverse susceptibility versus temperature, with a Curie-Weiss-type fit (red line) and extracted parameters.}
    \label{fig:Chi}
\end{figure}

As shown in Fig.~\ref{fig:Chi}(a), there is no sign of ordered magnetism down to 2~K. The magnetic susceptibility as a function of temperature is featureless, and a low temperature magnetization curve up to $\pm$7~T showed no significant hysteresis, allowing us to put an upper limit of about 0.2~T on any potential coercive field. This small value could be attributed to free cobalt, as no transitions to an ordered state were seen at or above this temperature. The data take on a Curie-Weiss form at low temperatures, and the inverse susceptibility curve could be fit to a straight line below 100~K [Fig.~\ref{fig:Chi}(b)]. We use the equation $\chi{} = \frac{C}{T - \theta{}_{CW}} + \chi{}_0$, with $C$ the Curie-Weiss constant, $\theta{}_{CW}$ the Curie-Weiss temperature, and $\chi{}_0$ the zero temperature susceptibility. We obtain $\theta_{CW} = -3.94 \pm 0.05$~K, which indicates antiferromagnetic interactions. This is further supported by a small effective moment (calculated from $C$) of just $\mu_{eff} = 0.036 \pm 0.003~\mu_{B}$ ($\mu_B$ is a Bohr magneton). No difference was observed between zero field-cooled and field-cooled data. Higher temperature susceptibility measurements up to 350~K continued the trend of Fig.~\ref{fig:Chi}. Given the lack of observed transition, small $\theta{}_{CW}$, lack of hysteresis, and extremely small moment, we believe we can rule out any long range order in KCo$_2$As$_2$, including at temperatures higher than those we measured. Alternate phases may have been the source of low temperature ferromagnetism identified in previous work \cite{AryalKCo2As2}, and powders would be much more susceptible to these sorts of impurities. Co$_2$As, for example, has a low temperature ferromagnetic transition \cite{ChenCo2As} at a temperature close to the $T_C$~=~15~K previously reported for KCo$_2$As$_2$.

\section{Electrical Transport and Heat Capacity}

\begin{figure*}
    \centering
    \includegraphics[width=\textwidth]{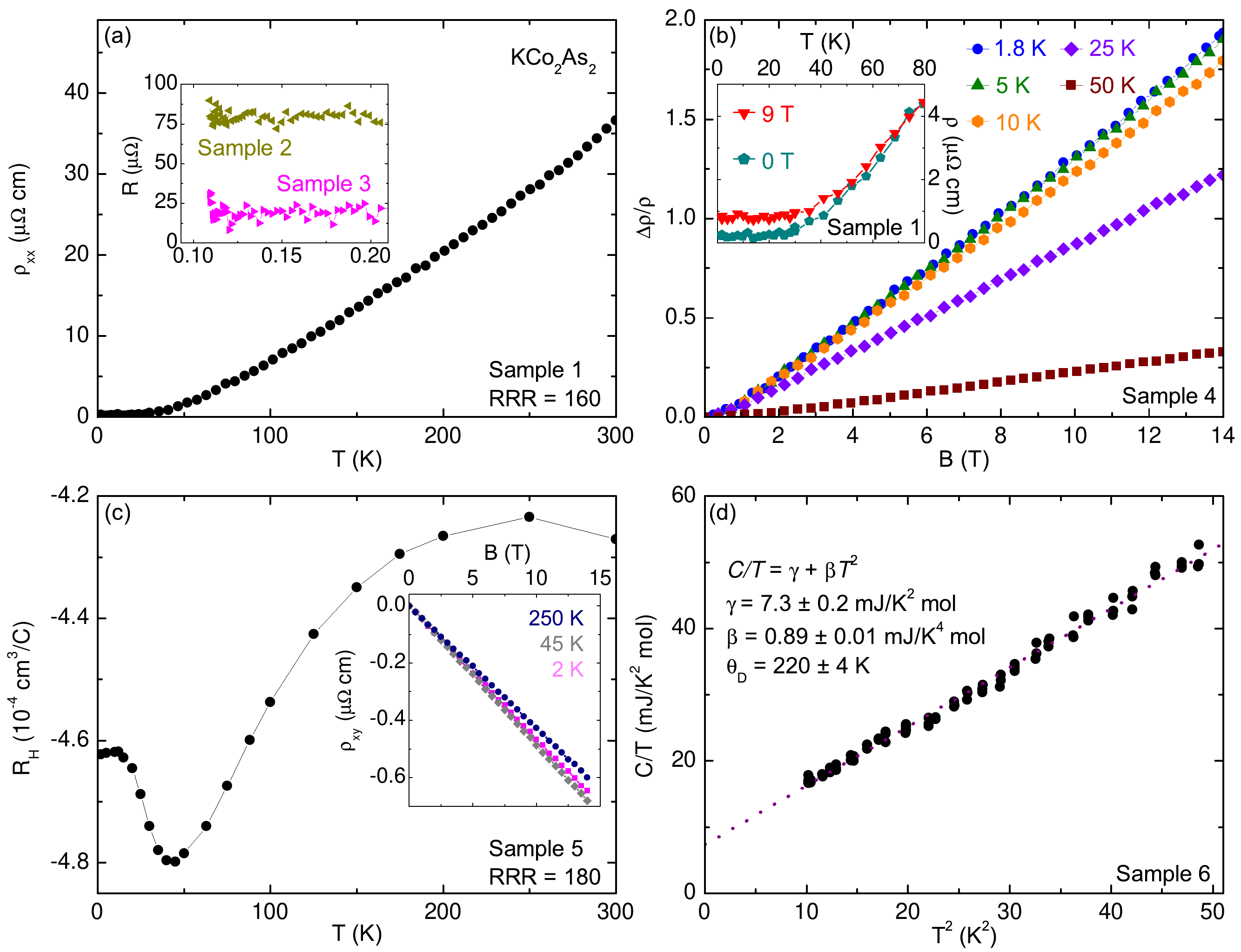}
    \caption{(a) Resistivity as a function of temperature in zero field with current in the \textit{ab}-plane. The inset shows the resistance of two other samples measured down to 100~mK, showing a lack of any low temperature transitions. (b) The magnetoresistance $\Delta{}\rho{}/\rho{} = \frac{\rho{}(B) - \rho{}(0~\textrm{T})}{\rho{}(0~\textrm{T})}$ of a different KCo$_2$As$_2$ sample at various temperatures, with the field again applied along the \textit{c}-axis perpendicular to current. The inset compares temperature sweeps at 0~T (the same data as (a)) and 9~T. (c) The Hall coefficient $R_H$ based on field sweeps at constant temperature. The inset shows the $\rho{}_{xy}$ data at 2, 45, and 250~K. All other field sweeps were also linear.  (d) The low temperature specific heat data. The purple dotted line is a linear fit based on the Debye low temperature model of the electron and phonon contribution.}
    \label{fig:Transport}
\end{figure*}

Electrical resistance was measured as a function of temperatures between 1.8 and 300~K and fields up to 14~T. Samples were shaped into rectangular bars with current in the \textit{ab} plane, using gold wires attached to the samples with silver paste. There were no distinct features in the resistivity as a function of temperature, which has an initial linear decrease from room temperature before becoming flat below about 30~K [Fig.~\ref{fig:Transport}(a)]. The shape is similar to a previous resistance measurement on pressed pellets \cite{ZinthKCo2As2}, but with a lower resistivity, which is to be expected for single crystals in comparison to polycrystals. The residual resistivity ratio, $\rho$(1.8~K)/$\rho$(300~K), is typically over 100 and as high as 300 for some crystals, indicating good crystal quality. Correspondingly, the 2~K resistivity can be as low as 0.25~$\mu\Omega$~cm, similar to the lowest values reported in the normal state for KFe$_2$As$_2$ \cite{LiuKFe2As2}, though that material has a higher $\rho{}$(300~K) and consequently larger RRR. Further measurements were made in zero field down to 100~mK with an adiabatic demagnetization refrigerator on multiple samples [Fig.~\ref{fig:Transport}(a),inset] There is no obvious change in resistance from $^4$He temperatures, i.e. no sign of superconductivity.

The magnetoresistance (MR) of KCo$_2$As$_2$, defined as $\frac{\rho(\textrm{B}) - \rho(0~\textrm{T})}{\rho(0~\textrm{T})}$ , becomes linear above about 1~T from the lowest temperatures up to 50~K [Fig.~\ref{fig:Transport}(b)], with little change in the crossover field with temperature. Measurements in negative field were symmetric, confirming that this was not the result of corruption from a Hall signal. The MR is about twice the zero field resistance at 1.8~K and 14~T and progressively smaller at higher temperature. Comparing temperature sweeps with and without field, MR only becomes sizeable below about 50~K [Fig.~\ref{fig:Transport}(b), inset].

Measurements of the Hall coefficient $R_H$ were made by sweeping field between -14~T and +14~T at constant temperature over the range 2-300~K [Fig.~\ref{fig:Transport}(c)]. Positive and negative field data were antisymmetrized to remove the longitudinal MR signal. The extracted even component was linear for all temperatures, as demonstrated by the Fig.~\ref{fig:Transport}(c) inset. The fitted slopes were used to calculate $R_H$. While the overall change is small, it is interesting to note the minimum near 50~K, before $R_H$ moves back toward zero as temperature is further decreased.

Heat capacity measurements were performed below 25~K on a 3.8~mg single crystal. Data from 3-7~K (10-50~K$^2$) fit well to the low temperature Debye specific heat model, $C/T = \gamma{} + \beta{}T^2$ [Fig.~\ref{fig:Transport}(d)]. The Sommerfeld coefficient  $\gamma = 7.3 \pm 0.2$~mJ/K$^2$~mol, and phonon term $\beta = 0.89 \pm 0.02$~mJ/K$^4$~mol can be used to solve for a Debye temperature of $\theta_D = 220 \pm 4$~K. These values, especially $\gamma$, are much closer to those of BaCo$_2$As$_2$ ($\gamma$~=~8.2~mJ/K$^2$~mol, $\theta_D$~=~250~K) \cite{SefatBaCo2As2} than KFe$_2$As$_2$ (90~mJ/K$^2$~mol and 160~K) \cite{KittakaKFe2As2}. The change in transition metal seems much more relevant to material properties than the change from alkali to alkaline earth, most notably in the correlation strength or effective mass, as indicated by $\gamma{}$. Combining the heat capacity and magnetization results, we can calculate the Wilson ratio $R_W = \frac{4\pi{}^{2}k_B{^2}\chi{}_0}{3\mu{}_0(g_e\mu{}_B)^{2}\gamma{}}$, where $\mu_0$ is the vacuum permeability, $k_B$ the Boltzmann constant, and $g_e$ the electron \textit{g}-factor. The closer a material is to long range magnetic order, the higher $R_W$ will be ($R_W = 1$ for a free electron gas). For KCo$_2$As$_2$ we get a value of about 2, much lower than BaCo$_2$As$_2$ where it is 7-10, depending on orientation \cite{SefatBaCo2As2}, reinforcing the weak magnetism seen in susceptibility data. It also shows that the substitution of alkali (K) for alkaline earth (Ba) does lead to some amount of change.
\section{ARPES}

\begin{figure*}[h!tb]
  \centering
  \includegraphics{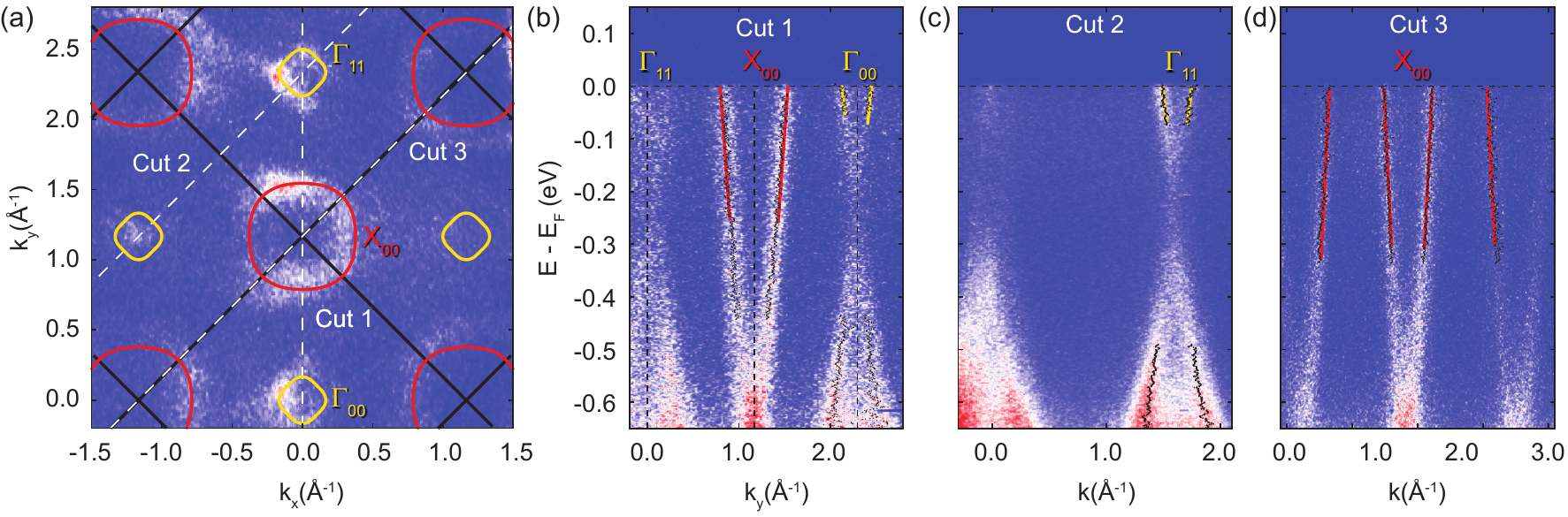}
  \caption{(a) The Fermi surface plotted as a false colour map, with the Brillouin zone indicated by solid black lines. The low-energy electronic structure is characterized by two electron-like pockets located at the zone center (yellow contour) and the zone corner (red contour). Note that the first BZ center ($k_x=k_y=0$) is at the bottom of panel (a), labelled as $\Gamma_{00}$. (b)-(d) The electronic dispersion along three different cuts [white dashed lines in (a)]. The quasiparticle dispersion (black markers) along the different cuts in momentum space is determined from fitting the momentum-distribution curves with Lorentzian line-shapes; the Fermi wave-vectors $k_\text{F}$ and velocities $v_\text{F}$ are determined from a linear fit of the quasiparticle dispersion, shown in yellow and red solid lines for the zone-center and zone-corner pockets, respectively.}
  \label{fig:arpes}
\end{figure*}

Angle-resolved photoemission spectroscopy (ARPES) measurements were carried out to quantitatively describe the KCo$_2$As$_2$ band structure and determine electron density. Single crystals were cleaved in the UHV environment at 11~K, exposing the plane perpendicular to the [001] direction. We performed photoemission using 120~eV photons in both linear-vertical and linear-horizontal polarizations, and summed the resulting ARPES intensity to produce the images in Fig.\,\ref{fig:arpes}. These measurements were taken with an R-4000 Scienta hemispherical analyzer at the Quantum Materials Spectroscopy Centre endstation of the Canadian Light Source. 

The Fermi surface [Fig.\,\ref{fig:arpes}(a)] is obtained by averaging the spectral intensity within a range of the Fermi energy ($E_F$) defined by the energy resolution ($\!\pm12$~meV), and consists of two pockets. The first (in yellow) is located at the Brillouin zone (BZ) center [$\mathbf{k^{\Gamma}} =\! (0, 0)$], and the second pocket (in red) is located at the BZ corner, $[\mathbf{k^X}\! =\! (\pm\!\pi/a, \pm\!\pi/a)]$. The BZ is indicated by solid black lines, using a value of $a\! =\! 3.809$~Å from the 150~K XRD results [Table~\ref{XRD150K}]. We take three cuts across the Fermi surface, shown in panels (b)-(d). From these ARPES spectra, we see that both the $\Gamma$ and $\text{X}$ pockets are electron-like. 

We extract the quasiparticle dispersion $E(\mathbf{k})$ from the momentum distribution curves (MDCs), which are given by the photoemission intensity at constant binding energy. As long as the self-energy is momentum-independent, the MDCs can be described by Lorentzian peaks centered at the quasiparticle momenta. We apply this Lorentzian fit (plus a constant background) to MDCs exhibiting sharp features; the extracted electronic dispersion $E(\mathbf{k})$ is shown as black markers in Fig.~\ref{fig:arpes}(b)-(d). Remarkably, $E(\mathbf{k})$ is well-described by a linear dispersion down to about 0.4~eV below the Fermi energy, as shown in yellow and red for the pockets at the BZ center and corner, respectively. We determine the Fermi velocity of the zone-corner pocket to be $v_\text{F}^{\text{X}}=2.4\pm0.3$~eV$\cdot$\AA~ from the linear fit of the bands shown in Fig.\,\ref{fig:arpes}(b) and (d), with no observed variation between the $\Gamma-\Gamma$ and $\Gamma-$X directions within uncertainty. Likewise, $v_\text{F}^\Gamma=1.8\pm0.2$~eV$\cdot$\AA~ is obtained from the $\Gamma_{00}$ and $\Gamma_{11}$ pocket in both high symmetry directions. The minimum of the $\text{X}_{00}$ pocket (extracted from a hyperbolic model of the dispersion) is located at $E_m^X=-0.62\pm0.06$~eV. However, this hyperbolic model underestimates the band minimum at $\Gamma_{00}$, so the energy-distribution curve is used in combination with a extrapolation of the linear dispersion to determine $E_m^\Gamma=-0.15 \pm 0.05$~eV. 

We note that density functional theory (DFT) calculations of the dispersion near the X point show the presence of two bands, while we observe only one: given that the dispersion of the two bands is very close ($\Delta k_\text{F}\approx 0.05~\text{\AA}^{-1}$), and matrix-element effects suppress the intensity of one of the bands in the high-symmetry cuts shown in Fig.\,\ref{fig:arpes}, we are not able to distinguish them within the broad features of the Fermi surface. Therefore, detailed modelling of the photoemission matrix elements and higher quality data is needed to confirm the dispersion of this second band.

To determine the electron concentration, we extract the Fermi wave-vector ($k_\text{F}$) of the electron pocket at $\Gamma$ and X. The tetragonal symmetry of the crystal lattice requires a four-fold symmetric in-plane Fermi surface. The simplest model satisfying this symmetry consists of a Fermi wave-vector with the following angle dependence,
\begin{equation}
\mathbf{k_\text{F}}-\mathbf{k^{\text{HS}}}=k_r+k_4\cos(4\phi),
\end{equation}
where $\phi$ is the angle along the Fermi surface with respect to the $\Gamma-\Gamma$ direction, $k_r$ and $k_4$ are parameters which define the radius and the warping of the Fermi surface, and $\mathbf{k^{\textbf{HS}}}$ is the momentum of the high-symmetry point at $\Gamma$ and $\text{K}$ for their respective electron pockets. The values for $k_r$ and $k_4$ are determined from the extracted $k_\text{F}$ values across the high symmetry directions $\Gamma-\Gamma$ and $\Gamma-\text{X}$.

We average the $k_{\text{F}}$ extracted from several cuts for each high-symmetry direction, and each pocket. $\text{X}_{00}$ is used for the analysis of the zone corner pocket. Along the $\Gamma-\text{X}$ direction, we find the Fermi wave-vector $k_\text{F}-\mathbf{k^\textbf{X}}=0.38 \pm0.02$~\AA$^{-1}$. Similarly, along the $\Gamma-\Gamma$ direction [Fig.\,\ref{fig:arpes}(d)], we obtain $k_\text{F}-\mathbf{k^\textbf{X}}=0.40\pm0.02$~\AA$^{-1}$. This gives $k_r=0.39 \pm0.03$~\AA$^{-1}$ and $k_4=0.012~\pm0.002$~\AA$^{-1}$, which results in an electron concentration of $0.35\pm0.05$ electrons per BZ (note that this electron concentration would be doubled by the presence of the additional band at X). The same analysis performed on the $\Gamma_{00}$ pocket at the zone center results in Fermi wave-vectors described by $k_\text{F}-\mathbf{k^\Gamma}=0.16 \pm 0.02$~\AA$^{-1}$ along the $\Gamma-\text{X}$ direction, and $0.14 \pm0.02$~\AA$^{-1}$ along the $\Gamma-\Gamma$ direction. This results in $k_r=0.15 \pm0.01$~\AA$^{-1}$ and $k_4=-0.010 \pm0.002$~\AA$^{-1}$, where the negative sign accounts for the 45 degree phase shift between the warping of the zone-center and zone-corner pockets. We calculate the electron concentration of the zone center pocket to be $0.052 \pm0.007$~electrons per BZ. When added together, the volume of all the Fermi surface sheets (including the additional pocket at X predicted by DFT, that we cannot distinguish) corresponds to a total carrier counting of about $0.75 \pm0.01$~electrons per BZ, or about 3.8~$\times{}~10^{21}$~ electrons per cm$^{3}$.

\section{Discussion}

The major themes of the iron pnictides are absent in KCo$_2$As$_2$. KFe$_2$As$_2$ shows clear evidence of strong correlations at low temperatures: it has a very large Sommerfeld coefficient, a steep drop in resistance at low temperature marking a coherent-incoherent crossover, and magnetic susceptibility reminiscent of heavy fermion compounds \cite{HardyKFe2As2}. The Co version instead features a $\gamma{}$ typical of metallic compounds, a resistance linearly proportional to temperature over most of the measured range, indicative of dominant electron-phonon scattering, and a small $\chi$ with Curie-Weiss-like low temperature behavior. The correlated behavior in KFe$_2$As$_2$ is attributed to coupling that places it close to a Mott transition \cite{HardyKFe2As2}. In fact, it has been shown that in general the six 3\textit{d} electrons result in the Fe pnictide parent compounds being close enough to half-filling of \textit{d} orbitals that correlation effects have significant influence, especially when chemical substitution or applied pressure are incorporated \cite{DeMediciMott}. Electron doping via Co moves it further away and, just as expected, signatures of correlated behavior vanish. Of course, as noted in analysis of band structure calculations, at the most basic level the addition of an electron changes the position of the Fermi energy and would have an effect even before considering correlations \cite{SinghKCo2As2}, as has been demonstrated already in comparisons of BaFe$_2$As$_2$ and BaCo$_2$As$_2$ \cite{XuBaCo2As2ARPES}. Still, while the loss of an electron through K-Ba substitution would seem to balance Co-Fe replacement, we do not see anything like the magnetic and structural transitions of BaFe$_2$As$_2$. This is because correlations in these compounds stem from the 3\textit{d} orbitals of the transition metal specifically, so doping on different sites does not cancel out. The presence of at least some iron is crucial, and cobalt cannot replicate its effects because its seventh electron disrupts the orbital order iron takes on, regardless of the other elements present.

A Fermi surface comparison again reveals parallels and key differences. In the nonmagnetic, tetragonal states of KFe$_2$As$_2$, BaFe$_2$As$_2$, and KCo$_2$As$_2$, ARPES data show pockets at the zone center and corner \cite{SatoKFe2As2ARPES,KondoBaFe2As2}. The first two in fact have several concentric pockets, with those at $\Gamma{}$ being of hole character. In contrast, KCo$_2$As$_2$ and (from previous studies) the \textit{AE}Co$_2$As$_2$ compounds \cite{Mao_ACo2As2,XuBaCo2As2ARPES} only have electron pockets. The K compound presents some differences even from the \textit{AE}Co$_2$As$_2$ series. Its $\Gamma{}$ pocket is circular, in contrast to the rhombus found at the zone center for the other materials \cite{Mao_ACo2As2}. This fits with calculations calling for cylindrical pockets \cite{SinghKCo2As2}.

The minimum in the Hall coefficient just below 50~K is unexpected for a single carrier type material. A ``turn on'' temperature in a similar range, featuring changes in both the MR and Hall coefficient, is seen in many extreme magnetoresistance materials, but typically involves changes in carrier concentration and mobility of multiple bands of differing sign \cite{TaftiLaSb}. ARPES also indicated that one band has a carrier concentration an order of magnitude higher, in which case it should dominate transport. However, the emergence of a significant MR and the change in zero field $\rho{}_{xx}$ in this same temperature range (from linear to roughly quadratic, before plateauing at even lower T) are signs for an overall change in scattering mechanism that would affect transport. It would not be detectable in our ARPES measurements since they were only done at a single, far lower temperature. Another interesting aspect is the linearity of the MR from relatively low field all the way up to 50~K, a departure from the expected quadratic dependence of a metal. There is also a difference between the carrier concentrations derived from ARPES and Hall effect measurements: for the former it is about 3.8~$\times{}~10^{21}$~cm$^{-3}$, while the latter is approximately 1.3~$\times{}~10^{22}$~cm$^{-3}$ at low temperatures. Whether this discrepancy is attributable to lack of experimental resolution or indicates something greater is a question to be resolved with continued study.

\section{Summary}

A reliable method for the production of air stable KCo$_2$As$_2$ single crystals has been provided and the properties from ambient to liquid helium temperatures summarized. This crystal growth technique opens up a route to potential synthesis of Rb- and Cs-based counterparts, where the incorporation of a heavier alkali metal would offer an interesting comparison. KCo$_2$As$_2$ does not exhibit any signatures akin to the highly correlated state of KFe$_2$As$_2$, and shows many differences from the general properties of other iron pnictide superconductors. This results from the extra electron of cobalt, which significantly alters 3$d$ orbital interactions and inhibits the same correlated effects seen in the iron equivalent, whose bands are closer to half filling. But KCo$_2$As$_2$ merits interest on its own. The observation of linear magnetoresistance and unusual hyperbolic dispersion of one of the electron bands in photoemission are signs that interesting physics might be found in this system. Future investigation of any changes around 50~K may reveal an instability to be exploited.

\section{Acknowledgments}

We thank Sabin Pokharel and Abdellah Lisfi at Morgan State University for assistance with magnetic susceptibility measurements. This work was supported by the National Science Foundation (NSF) Division of Materials Research award no.~DMR-1610349, the Gordon and Betty Moore Foundation's EPiQS Initiative through grant no.~GBMF9071, and the NSF Career award DMR-1455118. It was also undertaken thanks in part to funding from the Max Planck-UBC-UTokyo Centre for Quantum Materials and the Canada First Research Excellence Fund, Quantum Materials and Future Technologies Program. This project is also funded by the Killam, Alfred P. Sloan, and Natural Sciences and Engineering Research Council of Canada's (NSERCs) Steacie Memorial Fellowships (A.D.); the Alexander von Humboldt Fellowship (A.D.); the Canada Research Chairs Program (A.D.); NSERC, Canada Foundation for Innovation (CFI); British Columbia Knowledge Development Fund (BCKDF); and the CIFAR Quantum Materials Program. Part of the research described in this work was performed at the Canadian Light Source, a national research facility of the University of Saskatchewan, which is supported by CFI, NSERC, the National Research Council (NRC), the Canadian Institutes of Health Research (CIHR), the Government of Saskatchewan, and the University of Saskatchewan. D.J.C. acknowledges the support of the Anne G. Wylie Dissertation Fellowship.

\bibliography{KCo2As2Refs}

\end{document}